\documentclass[reprint, 
superscriptaddress, 
amsmath,
amssymb, 
aps,
floatfix, 
prb, 
]{revtex4-2}
\usepackage{graphicx} 
\usepackage{geometry}
\usepackage{color, xcolor}
\usepackage{ulem}
\usepackage{hyperref}

\newcommand{\br}{\mathbf{r}}
\newcommand{\bx}{\boldsymbol{r}}
\newcommand{\bX}{\boldsymbol{R}}

\newcommand{\bk}{\mathbf{k}}
\newcommand{\bq}{\mathbf{q}}
\newcommand{\bQ}{\mathbf{Q}}
\newcommand{\bP}{\mathbf{P}}

\newcommand{\bu}{\mathbf{u}}
\newcommand{\beps}{\boldsymbol{\varepsilon}}
\newcommand{\be}{\boldsymbol{e}}

\newcommand{\intd}{\mathrm{d}}

\newcommand{\NbSeFourHa}{\mathrm{4Ha}\textrm{\,--\,}\mathrm{NbSe}_2}
\newcommand{\NbSeTwoH}{\mathrm{2H}\textrm{\,--\,}\mathrm{NbSe}_2}

\date{\today}
\begin{document}
\title{Mapping Strain and Spatial-Modulation Phase from Local Wave Vectors in Scanning Tunneling Microscopy}

\author{Xinze Yang}
\thanks{These authors contributed equally to this work.}
\affiliation{Department of Physics, Yale University, New Haven, Connecticut 06511, USA}
\affiliation{Energy Science Institute, Yale University, West Haven, Connecticut 06516, USA}

\author{Sandra Sajan}
\thanks{These authors contributed equally to this work.}
\affiliation{Donostia International Physics Center, Donostia-San Sebasti\'an, Spain}

\author{Maria N. Gastiasoro}
\affiliation{Donostia International Physics Center, Donostia-San Sebasti\'an, Spain}

\author{Miguel M. Ugeda}
\affiliation{Donostia International Physics Center, Donostia-San Sebasti\'an, Spain}
\affiliation{Ikerbasque, Basque Foundation for Science, Bilbao, Spain}
\affiliation{Centro de F\'isica de Materiales CSIC-UPV/EHU, San Sebasti\'an, Spain}

\author{Eduardo~H.~da~Silva~Neto}
\email[Corresponding Author: ]{eduardo.dasilvaneto@yale.edu}
\affiliation{Department of Physics, Yale University, New Haven, Connecticut 06511, USA}
\affiliation{Energy Science Institute, Yale University, West Haven, Connecticut 06516, USA}
\affiliation{Department of Applied Physics, Yale University, New Haven, Connecticut 06511, USA}

\begin{abstract}
Spatially varying lattice distortions can produce local changes in electronic properties that are obscured by measurements averaging over extended regions. With its atomic-scale spatial resolution, scanning tunneling microscopy (STM) provides, in principle, a means of quantifying such distortions locally. A widely used approach for extracting local strain from STM images is the Lawler--Fujita (LF) algorithm [Lawler \textit{et al.}, Nature \textbf{466}, 347 (2010)], which determines the displacement field from the phase of the lattice modulation and subsequently obtains the strain by differentiation. However, this approach requires two-dimensional phase unwrapping, which can fail in regions where the phase is ill-defined, particularly near topological defects such as edge dislocations. Here, we introduce an alternative approach that circumvents these limitations by directly measuring the local wave vector in real space. We derive the relationship between the local wave vector, strain, and the phase gradient, and show that the local wave vector mapping (LWM) algorithm remains robust even when the phase itself is ill-defined. This makes the method particularly well suited for imaging the phase-gradient textures associated with vortices and other topological defects. We benchmark the algorithm using simulated data containing known lattice distortions and artificially generated vortices. Finally, we apply the method to experimental STM data and resolve the charge density wave (CDW) incommensurabilities, lattice distortions, and CDW phase vortices in $\NbSeFourHa$ and $\NbSeTwoH$. The physical implications of these observations were recently reported by S.~Sajan, X.~Yang, \textit{et al.} [arXiv:2607.20335 (2026)]. The program package and the example usage is publicly available at the GitHub: \href{https://github.com/jsntxyang/LocalWavevectorMeasuremnt}{https://github.com/jsntxyang/LocalWavevectorMeasuremnt}.
\end{abstract}

\maketitle
\section{Introduction}
The coupling between lattice structure and electronic behavior is a central theme in quantum materials \cite{Fradkin2010,Fernandes2014,Fernandes2022,Boehmer2022}. Even small lattice distortions can modify electronic properties by changing orbital overlap, local symmetry, or the balance among competing ordered states. One particularly clear manifestation of this coupling occurs in iron-based superconductors, where uniaxial strain couples strongly to electronic nematicity and can significantly modify the electronic structure \cite{Fernandes2014,Fernandes2022,Boehmer2022}. Such measurements establish strain as an important tuning parameter, but generally probe spatially averaged responses and therefore cannot determine how local lattice distortions affect electronic properties at the microscopic level. This distinction is particularly important because naturally occurring local strain can quench nematic fluctuations, which are thought to play an important role in quantum critical phenomena and superconductivity \cite{Kuo2016,Licciardello2019,Worasaran2021,Nag2024a}. More broadly, strong coupling between strain and electronic behavior has been observed across diverse classes of quantum materials, including cuprate, iron-based, and ruthenate superconductors, transition-metal dichalcogenides, and moiré materials, where lattice distortions can modify superconductivity, charge-density-wave and nematic order, dopant distributions, and electronic band structure \cite{Hicks2014,Zeljkovic2014,Gao2018,Kim2018,Boyle2021,Zhao2021b,Benschop2021,Mesple2021,Gupta2023,Du2023,Sajan2026}.

Over the past several decades, scanning tunneling microscopy (STM) has become an established technique for investigating quantum materials locally \cite{Yazdani2016, Rhodes2026}. By combining topographic imaging with spectroscopic measurements, STM can characterize lattice structure and electronic properties on the same atomic length scale. It has therefore played an important role in studying how local structural variations, including lattice strain, correlate with electronic inhomogeneity in quantum materials \cite{Zeljkovic2014, Gao2018, Pasztor2019, Zhao2021b, Benschop2021, Mesple2021, Du2023, Sajan2026, Niggli2026}. Over roughly the past fifteen years, a widely used method for extracting local lattice distortions from STM images has been to apply the Lawler--Fujita (LF) algorithm \cite{Lawler2010}. The LF algorithm was originally developed to correct image distortions caused by instrumental drift, such as piezoelectric creep and thermal drift. It determines a displacement field, relative to an idealized undistorted lattice, from the real-space phase of the lattice modulation. The image can then be corrected by removing this displacement field. Since instrumental drift typically produces smoothly varying distortions over fields of view of tens of nanometers, its contribution can often be estimated by fitting a smooth background. The residual displacement field, obtained after subtracting this smooth background, can then be interpreted as arising from local physical distortions of the lattice. From this residual field, strain can be obtained by taking spatial derivatives \cite{Benschop2021, Zhao2021b}. However, this approach has two important limitations for strain mapping. First, because strain is obtained by differentiating the displacement field, experimental noise can be amplified. Second, as discussed in Sec.~\ref{Section: Advantage}, the lattice phase can become ill-defined when a topological defect appears within the field of view (FOV). In this case, the displacement field obtained from the LF algorithm contains a nonphysical branch cut that cannot be removed by conventional phase unwrapping. Together, these limitations motivate methods that determine strain without first requiring reconstruction and differentiation of a displacement field.

In this work, we describe a new method that extracts local strain directly from STM images, rather than requiring phase unwrapping of a lattice modulation to reconstruct a displacement field that must then be differentiated to obtain the strain. Our method is based on the fact that a local distortion of the real-space lattice is equivalently described by a local change in the Bragg vectors. Thus, local strain can be obtained by measuring how the Bragg vectors vary across the image. A natural approach would be to use local Fourier transforms \cite{Jeon2024}, but this strategy has an intrinsic limitation: reducing the real-space field of view improves spatial resolution while simultaneously degrading the resolution in Fourier, or $\bq$, space. To overcome this tradeoff, we introduce a method based on the line Hough transform, a computer-vision algorithm that identifies straight-line features by mapping them into a parameter space. This allows us to determine Bragg vectors directly from real-space lattice features rather than from a locally windowed Fourier transform.

\section{Local Wave Vector Measurement (LWM) Algorithm}

\subsection{Relation between Strain and Local Bragg Vector}

\begin{figure}[!ht]
    \centering
    \includegraphics[width=1\linewidth]{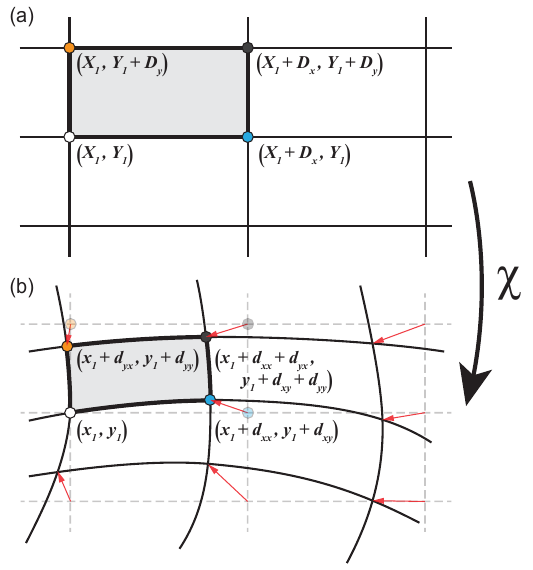}
    \caption{(a-b) An illustration showing how the deformation map, i.e., the $\chi$ map, changes the original grid $(X, Y)$ (a) to a deformed grid $(x, y)$ (b). Red vectors in (b) are the displacement field $\bu(\bX)$. $D_{x, y}$ define a rectangular grid unit in the undeformed $(X, Y)$  coordinate while the $d_{ij}, i, j \in \{x, y\}$ define a grid unit after the deformation. Since the deformation can be uniaxial and shear simultaneously, the new grid has both uniaxial components ($d_{xx}, d_{yy}$) and shear components ($d_{xy}, d_{yx}$).  The strain field is the derivative of the displacement field. In the system illustrated here, different strain components can be approximately expressed as $\beps_{ij} \approx \frac{d_{ij}}{D_i} - \delta_{ij} (i, j \in \{x, y\})$.}
    \label{fig: Chi Map}
\end{figure}

Strain provides a local description of the deformation of a body. Assuming that a body occupies a region $B$ before deformation and $B'$ after deformation, a one-to-one map $\chi: B \to B'$ can be established between the original body $B$ and the deformed body $B'$ (see Fig.\,\ref{fig: Chi Map}): 
\begin{equation}
\bx = \chi(\bX)
\end{equation}
Here, $\bX$ denotes the coordinates of points in the undeformed body $B$, Fig.\,\ref{fig: Chi Map}(a), and $\bx$ denotes the corresponding coordinates of the same $\bX$ points in the deformed body $B'$, Fig.\,\ref{fig: Chi Map}(b). By definition, the unsymmetrized strain tensor, which is equivalent to the displacement-gradient tensor, is given by the derivative of the displacement field $\bx - \bX$ with respect to $\bX$, i.e., 
\begin{equation}
\label{eq: Strain Definition}
\beps(\bX) = \nabla \chi(\bX) - I,
\end{equation}
with the identity $I$. The corresponding standard symmetric strain tensor commonly used in the literature is $(\beps^T + \beps)/2$.

In the $\bx$ coordinates, an STM topographic image can be represented as a map $\tilde{\mathcal{S}}: B' \to \mathbb{R}$, where $\tilde{\mathcal{S}}(\bx)$ denotes the measured topographic signal at position $\bx$. Even when the underlying lattice is perfectly periodic, the measured image generally does not display a perfectly periodic structure because of instrumental drift. These distortions broaden the Bragg peaks obtained from the Fourier transform of $\tilde{\mathcal{S}}(\bx)$. In contrast, when the same lattice is described in the undeformed coordinates $\bX$, the corresponding reference map $\mathcal{S}: B \to \mathbb{R}$ can be expanded in Fourier components located at singular, well-defined reciprocal lattice vectors:
\begin{equation}
\label{eq: STM image undeformed}
\mathcal{S}(\bX) = \sum_{\bQ} A_{\bQ} e^{i\bQ^T\bX},
\end{equation}
where $\bQ = (Q_X, Q_Y)^T$ are reciprocal lattice vectors and $A_{\bQ}$ are the associated Fourier coefficients. 
Throughout this paper, the inner product of column vectors is expressed as matrix multiplication, such that $\bQ^T\bX=Q_X X+Q_Y Y$ for $\bQ=(Q_X,Q_Y)^T$ and $\bX=(X,Y)^T$, as in Eq.\,(\ref{eq: STM image undeformed}).
By definition of the deformation mapping $\chi$, the measured image $\tilde{\mathcal{S}}(\bx)$ and the reference map $\mathcal{S}(\bX)$ are related by
\begin{align}
\mathcal{\tilde{S}}(\bx) &= \mathcal{S} ( \chi^{-1}(\bx)) \nonumber \\
                         &= \sum_{\bQ} A_{\bQ} e^{i\bQ^T\chi^{-1}(\bx)} .
\label{Function in Euler Coordinate}
\end{align}
In STM experiments, the available information consists of the distorted STM topographic image, i.e., $\tilde{\mathcal{S}}(\bx)$, together with the Bragg vectors $\bQ$ of the reference lattice. The central problem is therefore to extract a map of $\nabla \chi(\bX)$ from $\tilde{\mathcal{S}}(\bx)$ and the known Bragg vectors $\bQ$. These Bragg vectors may be obtained either directly from the STM image by Fourier transform or from independent experimental techniques such as X-ray diffraction (XRD). In the latter case, the reference values of $\bQ$ may differ from the Bragg vectors observed in the FFT of the STM image. Such a discrepancy contributes only a constant background to the strain map and does not affect the analysis of relative strain variations.

\begin{figure*}
    \centering
    \includegraphics[width=0.85\linewidth]{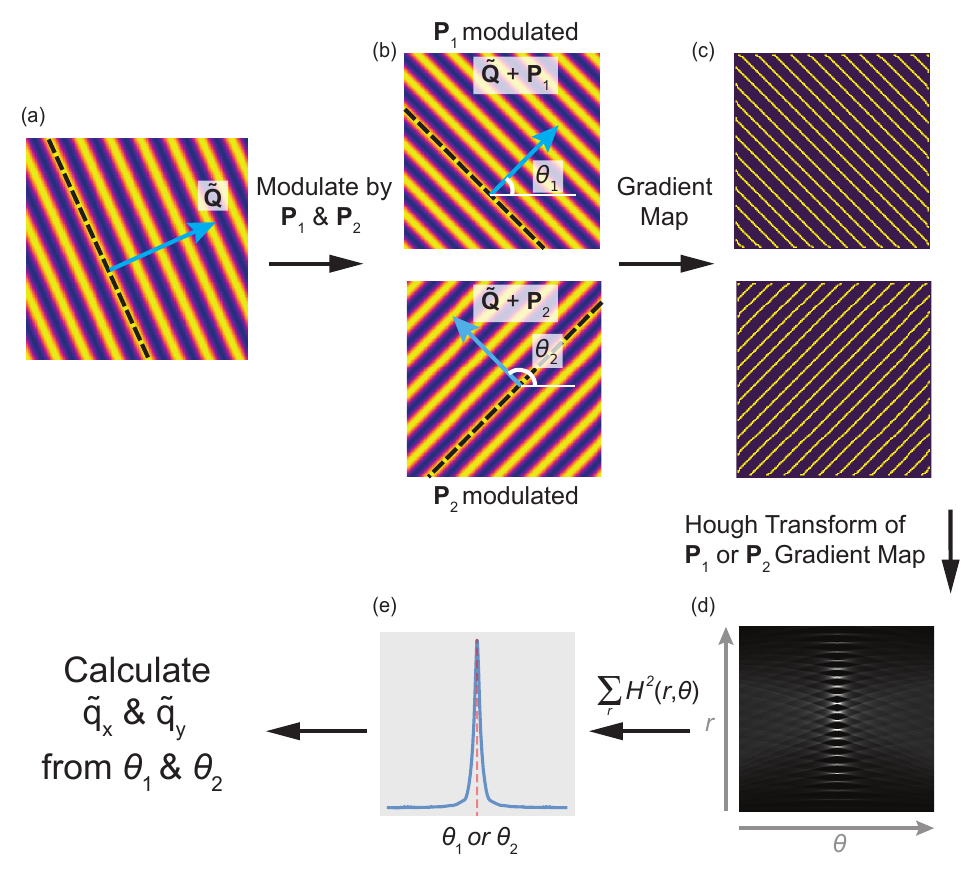}
    \caption{(a-e) Illustrations showing the process for measuring the local wave vector using the Hough transform.}
    \label{fig: Local Bragg Process}
\end{figure*}

Using the definitions above, we can relate the deformation encoded in the STM image to the Bragg vectors measured locally in that image. Assuming that $\chi(\bX)$ is sufficiently smooth and varies slowly on the scale of the lattice spacing, it can be expanded to first order around a point $\bX_0$:
\begin{equation}
\label{eq: Local Expansion}
\bx - \bx_0 \approx \nabla \chi(\bX_0) (\bX - \bX_0),
\end{equation}
which can be re-written as
\begin{equation}
\label{eq: Local Expansion 2}
\bX = \bX_0 + [\nabla \chi(\bX_0)]^{-1}(\bx - \bx_0).
\end{equation}
Using Eq.\,(\ref{eq: Strain Definition}), Eq.\,(\ref{eq: Local Expansion 2}) becomes 
\begin{equation}
    \bX = \bX_0+ (I + \beps)^{-1}(\bx - \bx_0) ,
\end{equation}
which can be used to substitute $\chi^{-1}(\bx) = \bX$ into 
Eq.\,(\ref{Function in Euler Coordinate}) to yield
\begin{align}
\mathcal{\tilde{S}}(\bx) & = \sum_{\bQ} A_{\bQ} e^{i\bQ^T \left[\bX_0 +(I + \beps)^{-1}(\bx - \bx_0) \right]} \\
                         & = \sum_{\bQ} A_{\bQ} e^{i\bQ^T\bX_0}e^{i[\bQ^T(I + \beps)^{-1}](\bx - \bx_0)} .
\end{align}
This expression shows that locally, in the vicinity of $\bX_0$, the STM image $\tilde{S}(\bx)$ modulates with 
a wave vector $\tilde{\bQ}(\bX_0)$  dependent on the strain $\beps(\bX_0)$ and the Bragg vector $\bQ$ of the reference lattice: 
\begin{equation}
\label{Eq. Bragg Strain Relation}
\tilde{\bQ}(\bX_0)^T = \bQ^T (I + \beps(\bX_0))^{-1}.
\end{equation}
Since $\bx_0$ and $\bX_0$ have a one-to-one correspondence, the expression in $\bx$, i.e., in the coordinate of the STM image, is exactly the same:
\begin{equation}
\label{Eq. Bragg Strain Relation in x}
\tilde{\bQ}(\bx_0)^T = \bQ^T (I + \beps(\bx_0))^{-1}.
\end{equation}

Equation\,(\ref{Eq. Bragg Strain Relation in x}) may be viewed as the reciprocal-space form of the Cauchy--Born rule \cite{Born1996}. It reduces the problem of measuring strain to that of measuring local wave vectors. A natural approach would be to perform a local fast Fourier transform (FFT) \cite{Jeon2024}. However, this approach is impractical for high-spatial-resolution strain mapping because reducing the real-space window size necessarily degrades the resolution in $\bq$ space. As discussed in the next subsection, we instead determine the local Bragg vectors directly from the real-space lattice modulations.

\subsection{Measuring Local Wave Vectors Using the Hough Transform}

To obtain the four components of $\beps(\bx)$ we need to solve Eq.\,(\ref{Eq. Bragg Strain Relation in x}) for at least two linearly independent $\tilde{\bQ}$ lattice vectors. We analyze each $\tilde{\bQ}$ separately. Given an STM image, a single plane wave image can be obtained by filtering a specific $\tilde{\bQ}$ vector in Fourier space. Due to the strain throughout the image, $\tilde{\bQ}$ is expected to vary in space. Still, with an appropriate choice of the filter window, a plane wave with distorted wave front can still be obtained. We then modulate the distorted wave front image ($\tilde{\mathcal{S}}(\bx)=\Re[e^{i\tilde{\bQ}^T\bx}]$, Fig.\,\ref{fig: Local Bragg Process}(a)) by two pre-selected plane waves, $e^{i\bP_1^T\bx}$ and $e^{i\bP_2^T\bx}$. After the modulation, the corresponding wave vectors are shifted to $\tilde{\bQ}_1 = \tilde{\bQ} + \bP_1$ and $\tilde{\bQ}_2 = \tilde{\bQ} + \bP_2$, respectively (Fig.\,\ref{fig: Local Bragg Process}(b)). By measuring the wavefront orientations in these two modulated images (\textit{i.e.}, $\theta_1$ and $\theta_2$ marked in Fig.\,\ref{fig: Local Bragg Process}(b)), the directions of $\tilde{\bQ}_1$ and $\tilde{\bQ}_2$ can be determined. This yields two equations:
\begin{align}
\frac{\tilde{q}_y + p_{1y}}{\tilde{q}_x + p_{1x}} &= \frac{\sin\theta_1}{\cos\theta_1}, \label{eq: Angle Relation 1} \\
\frac{\tilde{q}_y + p_{2y}}{\tilde{q}_x + p_{2x}} &= \frac{\sin\theta_2}{\cos\theta_2}. \label{eq: Angle Relation 2}
\end{align}
Since $\bP_1$ and $\bP_2$ can be chosen such that Eqs.\,\ref{eq: Angle Relation 1} and \ref{eq: Angle Relation 2} are nondegenerate, the values of $\tilde{q}_x$ and $\tilde{q}_y$ can be uniquely determined. The original wave vector $\bQ$ can thus be reconstructed from the spatial map of $\theta_1$ and $\theta_2$.

To determine the wavefront angles we first convert each modulated intensity image into a gradient map, shown in Fig.\,\ref{fig: Local Bragg Process}(c) \cite{Canny1986}. We then apply the Hough transform (HT), a standard line-extraction technique widely used in computer vision \cite{Duda1972} to obtain $\tilde{\bQ}_1$ and $\tilde{\bQ}_2$ from the gradient images. In two dimensions, a line can be parameterized by its distance $r$ from the image center and by an angle $\theta$ specifying its orientation. The Hough transform maps an image containing lines into a distribution over the parameters $(r,\theta)$, which we denote as $H(r,\theta)$. Peaks in this distribution correspond to the parameters of the lines present in the image. As shown in Fig.\,\ref{fig: Local Bragg Process}(d), the HT maps the parallel wavefronts into bright features concentrated at the same $\theta$ value. Once $H(r,\theta)$ is obtained, we integrate $H^2(r,\theta)$ along the $r$ direction:
\begin{equation} 
h(\theta) = \int H^2(r,\theta)\,\intd r\,, 
\end{equation} 
where the dominant peak in $h(\theta)$ corresponds to the wavefront angle, as illustrated in Fig.\,\ref{fig: Local Bragg Process}(e). 
In our algorithm, the location of the peak $h(\theta)$ is obtained from fitting it to a Gaussian function. 
This procedure is applied separately to the two modulated images, corresponding to the $\bP_1$ and $\bP_2$ shifts, yielding the two angles $\theta_1$ and $\theta_2$ used in Eqs.\,\eqref{eq: Angle Relation 1} and \eqref{eq: Angle Relation 2}. The full procedure is illustrated in Fig.\,\ref{fig: Local Bragg Process}. This process determines the wave vector $\tilde{\bQ}$ at a single point corresponding to the center of the analysis window. To construct a spatially resolved local wave vector map, a small window is scanned across the entire image, and the local wave vector is measured at each position.

\subsection{Advantages of the LWM Algorithm} \label{Section: Advantage} 

The key difference between the LWM and LF algorithms lies in the quantity extracted directly from the STM image. In the LF algorithm, the spatially varying phase of a lattice modulation is first extracted and unwrapped. A component of the displacement field is then inferred from this phase, and the corresponding strain is obtained by differentiation. 

In the LF algorithm, the displacement field $\bu(\bx)$ appears in the expression describing the lattice modulation:
\begin{equation}
I(\bx) = I_0 \Re\left[e^{i[\bQ^T\bx + \phi(\bx)]}\right],
\end{equation}
where the modulation phase is
\begin{equation}
\phi(\bx) = -\bQ^T\bu(\bx).
\end{equation}
Multiplying the image by $e^{-i\bQ^T\bx}$ removes the lattice contribution, resulting in the phase image $e^{i\phi(\bx)}$. However, the complex exponential is defined modulo $2\pi$, and extracting its argument results in a wrapped phase in the interval $(0,2\pi)$, or equivalently $(-\pi,\pi)$. If the phase can be correctly unwrapped, a continuous representation of $\phi(\bx)$ extending beyond the principal interval can be obtained, and thus the projection $\bQ^T\bu(\bx)$ of the displacement field can be recovered. This shows that the LF algorithm fundamentally relies on phase extraction and unwrapping.

In contrast, the LWM algorithm determines the local deformation gradient directly, without the need to first construct a phase map or a displacement field. In a small region centered at $\bx_0$ where the phase term $\phi(\bx)$ can be approximated by its first-order expansion,
\begin{equation}
I(\bx) = I_0 \Re\left[e^{i[\bQ^T\bx + \phi(\bx_0) + (\nabla\phi(\bx_0))^T (\bx-\bx_0)]}\right]
\end{equation}
This shows that the locally observed wave vector and the phase term have the following relation.
\begin{equation}
\label{eq: Local Wave Vector and Phase Relation}
\tilde{\bQ}(\bx) = \bQ_0 + \nabla\phi(\bx)
\end{equation}

When the phase of the lattice modulation can be unwrapped reliably, the LF algorithm provides an adequate method for constructing displacement and strain maps, provided that noise amplification from numerical differentiation remains manageable. In the presence of a topological defect, however, the phase, and consequently the displacement, cannot be represented as a globally continuous, single-valued field, preventing its derivative from yielding a well-defined strain field throughout the image.

\begin{figure}[!ht]
    \centering
    \includegraphics[width=1.0\linewidth]{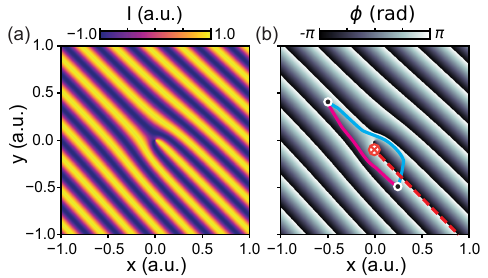}
    \caption{(a) Plane wave with a topological defect at the center. (b) The wrapped phase of the plane wave shown in (a). The magenta and blue curves shows two difference path along which the phase jumps by 0 and $2\pi$ respectively. The red $\otimes$ symbol marks the center of the vortex and the read dashed line marks the new wave front branch indeuced by the vortex.}
    \label{fig: Phase Definition}
\end{figure}

A phase vortex provides a simple example of this topological obstruction. In a crystal lattice, such a defect can arise from an edge dislocation, where the termination of a lattice row produces a phase winding in the lattice modulation. Similar singularities can occur in charge-density-wave (CDW) order, for example near impurities or CDW domain boundaries. A minimal model is given by the distorted plane wave \begin{equation} \label{eq: Vortex Function} I(\bx) = \Re\{e^{i(\bk^T\bx + \arg(\bx))}\}, \end{equation} where $\arg(\bx)=\arg(x+iy)$ is the polar angle of the vector $\br$, with the branch cut chosen along the $-y$ axis. The resulting image is shown in Fig.\,\ref{fig: Phase Definition}(a), where an additional wavefront terminates at the center of the image. This termination corresponds to the phase vortex marked by the red $\otimes$ in the wrapped phase map in Fig.\,\ref{fig: Phase Definition}(b). Away from the vortex, the discontinuities in the wrapped phase are ordinary $2\pi$ wrapping discontinuities and can be removed locally by adding appropriate multiples of $2\pi$. Around the vortex, however, the phase accumulates a net winding of $2\pi$ along any closed loop enclosing the defect, such as the loop formed by the magenta and blue paths in Fig.\,\ref{fig: Phase Definition}(b). Near this defect, continuously unwrapping the phase between the the starting and end points marked in the figure would assign values differing by $2\pi$ to the same end point, depending on whether the blue or magenta path is chosen. A globally continuous, single-valued phase field therefore cannot be constructed, and a discontinuity must remain along a branch cut, indicated by the red dashed line in Fig.\,\ref{fig: Phase Definition}(b). Although a different choice of phase representation can move the branch cut, it cannot eliminate it. The displacement field inferred by the LF algorithm consequently contains the same nonphysical discontinuity, which produces an artificial contribution when the field is differentiated to obtain strain. The LWM algorithm avoids this problem because it does not rely on constructing a phase field. Synthetic and experimental examples involving phase vortices are presented in Secs.\,\ref{subsection: Topological Defects} and \ref{section: Example Local Wave Vector Map}, respectively.

\section{Benchmarking with Synthetic Data} 
\label{section: Synthetic Data}

\subsection{Example 1: Drift Correction with a Known Strain Field} 
\label{subsection: Drift Correction}

\begin{figure}[!htbp]
    \centering
    \includegraphics[width=1.0\linewidth]{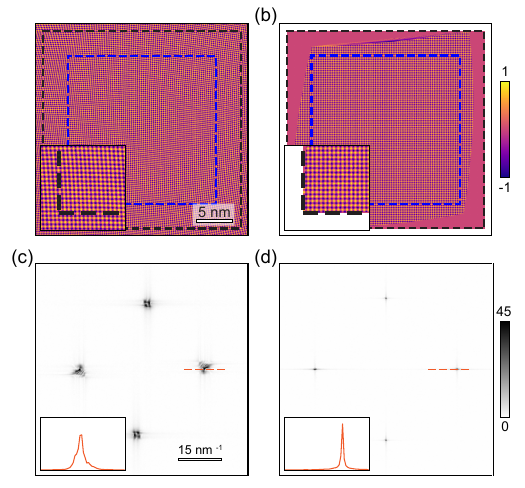}
    \caption{(a,b) Distorted lattice before and after distortion correction, respectively. The insets show magnified views of the lower-left corners, and the scale bar in (a) represents $5\mathrm{nm}$. The black dashed squares indicate the regions where the LWM algorithm is applied, while the blue dashed squares indicate the regions used to calculate the FFTs shown in (c,d). (c,d) FFTs of the regions marked in (a,b), respectively; the insets show line cuts taken along the orange dashed lines.}
    \label{fig: Correction}
\end{figure}

We first test the ability of the LWM algorithm to correct drift and recover a known strain field from an STM-like image. Because the true strain field is not generally known in experimental STM data, we generate a synthetic image with a prescribed displacement field, as shown in Fig.\,\ref{fig: Correction}. The image is defined as 
\begin{equation} 
I(\bx) = \Re\{e^{i\bk^T (\bx - \bu(\bx))}\} + \mathrm{noise}, 
\end{equation} where $\bu(\bx)$ describes the imposed distortion in the image coordinates and the noise is added pixel-wise with a uniform distribution between $[-0.02, 0.02] \times \mathrm{Lattice\ Amplitude}$. We then apply the LWM algorithm to measure the strain field and compare the result with the known gradient of $\bu(\bx)$, 
\begin{equation}
    \be_{ij} = \frac{\partial \bu_{i}}{\partial \bx_j}. 
\end{equation} 
It is important to note that $\be_{ij}$ is not identical to the strain tensor $\beps_{ij}$ defined in Eq.\,(\ref{eq: Strain Definition}), because the derivative in $\be_{ij}$ is taken with respect to the deformed coordinate $\bx$, rather than the undeformed coordinate $\bX$. Instead, the two quantities are related by 
\begin{equation} 
    \label{eq: Body to Lab}
    \be = (I + \beps)^{-1}\beps . 
\end{equation} 
For this reason, we use the LWM algorithm to obtain the strain tensor $\beps$ and then calculate $(I + \beps)^{-1}\beps$ for comparison with the known field $\be$.

To simulate realistic STM data, $\bu(\bx)$ is composed of two terms: 
\begin{equation} 
\bu(\bx) = \bu_{\mathrm{dr}}(\bx) + \bu_{\mathrm{ph}}(\bx). 
\end{equation} 
The first term, $\bu_{\mathrm{dr}}(\bx)$, is taken to be a two-dimensional second-order polynomial, which simulates the drift induced during the STM scan. The second term, $\bu_{\mathrm{ph}}(\bx)$, is generated by Gaussian smoothing a randomly sampled binary matrix, and is used to simulate local physical strain. This simulates a situation in which the physical strain varies over shorter length scales than the instrumental drift distortion, with the drift component changing smoothly and approximately monotonically across the image while the physical strain fluctuates locally around a background value. Thus, previous STM strain-mapping studies \cite{Zhao2021b} have isolated $\bu_\mathrm{ph}(\bx)$ by subtracting a low-order polynomial background from $\bu(\bx)$. Likewise, we subtract a fitted second-order polynomial background from the total extracted image strain.

\begin{figure*}[!tbp]
    \centering
    \includegraphics[width=0.99\linewidth]{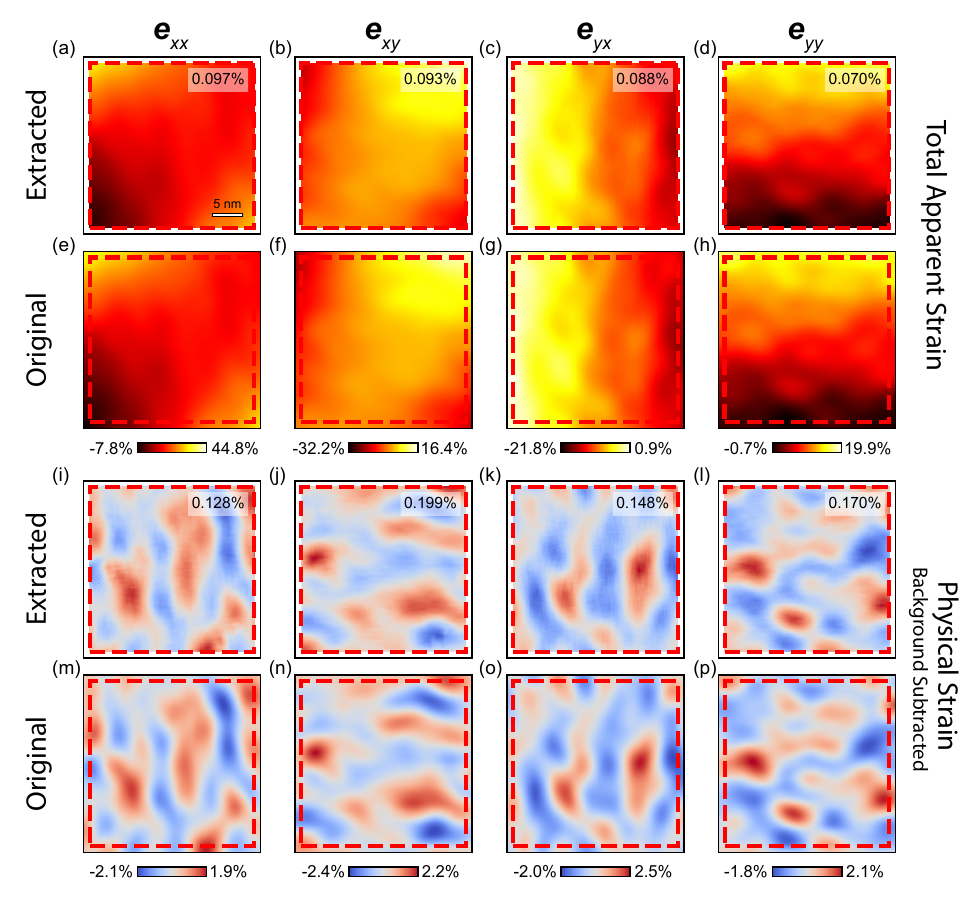}
    \caption{(a--h) Comparison between the original and extracted total apparent strain maps, including contributions from instrumental drift. (i--p) Corresponding comparison after background subtraction. Values in the upper-right corners of the extracted maps indicate the RMS errors defined in Eq.\,(\ref{eq: RMS}).}
    \label{fig: Fig Strain Comparison}
\end{figure*}

The maps of the total distortion gradient and the physical distortion gradient obtained using the LWM algorithm are shown in Fig.\,\ref{fig: Fig Strain Comparison}. To characterize the accuracy of the LWM algorithm, we computed the root-mean-square error (RMSE) between the measured value and the correct value (numbers on the upper right corner of Figs.\,\ref{fig: Fig Strain Comparison}(a-d) and (i-l)):
\begin{equation}
\mathrm{RMSE}_{\,ij} = \sqrt{\frac{1}{N}\sum_{\bx}([\be_{\mathrm{ori}}(\bx)]_{ij}-[\be_{\mathrm{ext}}(\bx)]_{ij})^2}
\label{eq: RMS}
\end{equation}
Here, $\be_{\mathrm{ori}}$ and $\be_{\mathrm{ext}}$ denote the original (correct) and LWM-extracted strain tensors, respectively, while $\sum_{\bx}$ denotes the sum over all $N$ pixels in the strain map. The values of $\be_{ij}$ predicted by the LWM algorithm agree well with the imposed distortion field, with RMSE below $1.0\times10^{-3}$ for all four components (Figs.\,\ref{fig: Fig Strain Comparison}(a-d)). The four components of the physical distortion gradient extracted after polynomial subtraction also agree quantitatively with the true values (Figs.\,\ref{fig: Fig Strain Comparison}(i-l)). The errors in this case are slightly larger than those obtained for the total distortion gradient, increasing to a root-mean-square value of approximately $1.5\times10^{-3}$, likely due to the polynomial fitting and subtraction procedure.

In addition to measuring the strain, the LWM algorithm can also be used to correct the distortion in the image. To perform this correction, one first reconstructs the displacement field by integrating the strain field:
\begin{align}
    \bu(\bx_1) & = \bu(\bx_0) + \int_{\bX(\bx_0)}^{\bX(\bx_1)} \beps(\bX) \, \intd \bX \label{eq: disp integral line1}\\
               & = \bu(\bx_0) + \int_{\bx_0}^{\bx_1} \beps(\bx) [I + \beps(\bx)]^{-1}\,\intd \bx \label{eq: disp integral line2}
\end{align}
As in the distinction between $\beps$ and $\be$, the displacement field is naturally obtained by integrating the strain in the undeformed coordinate $\bX$. In practice, however, the strain map measured from the STM image is sampled on a regular grid in the deformed image coordinate $\bx$. It is therefore necessary to rewrite the integral in terms of $\bx$. Using Eq.\,(\ref{eq: Local Expansion}), the change of variables from $\intd\bX$ to $\intd\bx$ introduces the factor $[I+\beps(\bx)]^{-1}$ in the integrand, leading to Eq.\,(\ref{eq: disp integral line2}).
Once the displacement field $\bu(\bx)$ is obtained, each pixel in the measured image is mapped to its corresponding coordinate in the undeformed frame according to $\bX=\bx-\bu(\bx)$. Because the resulting coordinates generally form an irregular grid, the measured image is interpolated onto a uniformly spaced grid in $\bX$. The corrected real-space image and its FFT are shown in Fig.\,\ref{fig: Correction}; the regular lattice and sharp Bragg peaks observed after correction demonstrate that the distortion has been accurately removed.

\subsection{Example 2: Phase-Gradient Mapping Near Topological Defects} 
\label{subsection: Topological Defects}

\begin{figure}[!tbp]
    \centering
    \includegraphics[width=0.95\linewidth]{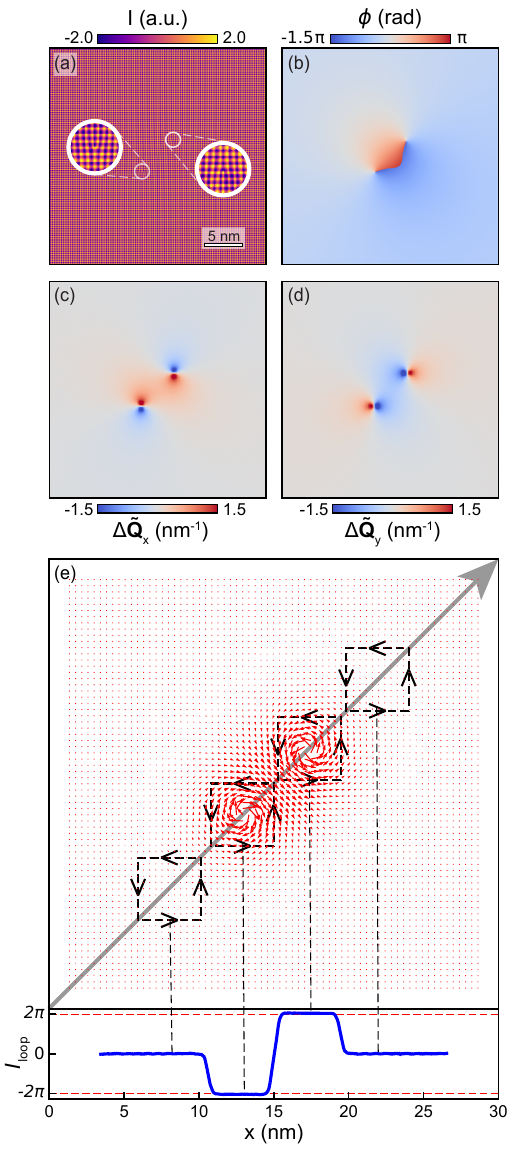}
    \caption{(a) Image of a lattice containing two vortices near the center; the insets show enlarged views of the individual vortices, and the scale bar represents $5$\,nm. (b) Unwrapped phase map obtained using the LF algorithm, with a branch cut connecting the two vortices. (c,d) Maps of the local components of $\Delta\tilde{\bQ}$. (e) Vector-field plot of $\Delta\tilde{\bQ}$; the bottom panel shows $I_{\mathrm{loop}}$ for loops centered at positions along the gray arrow in the upper panel.}
    \label{Fig: Algorithm Result}
\end{figure}

The second example demonstrates the ability of the LWM algorithm to measure phase gradients near topological defects, where the global phase is ill-defined and conventional phase-unwrapping procedures fail. To illustrate this capability, we generate a $30\,\mathrm{nm} \times 30\,\mathrm{nm}$ image of a two-dimensional square lattice with lattice constant $0.25\,\mathrm{nm}$. Using Eq.\,(\ref{eq: Vortex Function}), two topological defects are introduced near the center of the image, as shown in Fig.\,\ref{Fig: Algorithm Result}(a), and the resulting distortion is analyzed using both the LWM and LF algorithms, the latter of which relies on phase unwrapping.

As shown in Fig.,\ref{Fig: Algorithm Result}(b), the \textit{unwrapped} phase obtained using the LF algorithm contains a branch cut, which may follow any path connecting the two defects and across which its derivative exhibits a nonphysical discontinuity. The phase gradient, $(\partial\phi/\partial x,\partial\phi/\partial y)$, therefore provides a better-defined local measure of the vortices and can be obtained directly using the LWM algorithm through its relation to the local wave vector [Eq.\,(\ref{eq: Local Wave Vector and Phase Relation})]:
\begin{align}
\tilde{\bQ}_x(\bx) &= \frac{\partial \phi}{\partial x} + \bQ_{x}, \label{eq: Qx Phase Relation}\\
\tilde{\bQ}_y(\bx) &= \frac{\partial \phi}{\partial y} + \bQ_{y}. \label{eq: Qy Phase Relation}
\end{align}
To remove the $\bQ$ background, we subtract the spatial averages of $\tilde{\bQ}_x(\bx)$ and $\tilde{\bQ}_y(\bx)$, revealing the phase-gradient texture generated by the topological defects. The resulting average-subtracted quantities are denoted by $\Delta\tilde{\bQ}_x$ and $\Delta\tilde{\bQ}_y$, i.e., \begin{align} \Delta\tilde{\bQ}_x(\bx) &= \tilde{\bQ}_x(\bx) - \langle\tilde{\bQ}_x(\bx)\rangle = \frac{\partial\phi}{\partial x}, \\ \Delta\tilde{\bQ}_y(\bx) &= \tilde{\bQ}_y(\bx) - \langle\tilde{\bQ}_y(\bx)\rangle = \frac{\partial\phi}{\partial y}. \end{align} These quantities are illustrated by the color maps in Fig.\,\ref{Fig: Algorithm Result}(c,d) and the vector-field plot in Fig.\,\ref{Fig: Algorithm Result}(e), where two distinct vortices corresponding to the imposed topological defects are clearly visible. Furthermore, the circulation of the phase gradient around a closed loop enclosing a vortex is quantized: 
\begin{equation} \begin{aligned} I_{\mathrm{loop}} &= \oint \Delta\tilde{\bQ}_x\,\mathrm{d}x + \oint \Delta\tilde{\bQ}_y\,\mathrm{d}y = 2\pi n, \\ &\hspace{2em} n=0,\,\pm1,\,\ldots \end{aligned} \end{equation}
This quantization is confirmed by integrating the phase gradient around a square loop of size $5\times5\,\mathrm{nm}^2$ as it is translated along the diagonal direction of the image. The bottom panel of Fig.\,\ref{Fig: Algorithm Result}(e) shows the resulting circulation, which takes the quantized values $\pm2\pi$ when the loop encloses a single vortex.

\section{Application to Experimental STM Data}
\label{section: Example Local Wave Vector Map}
\begin{figure*}
    \centering    \includegraphics[width=0.98\linewidth]{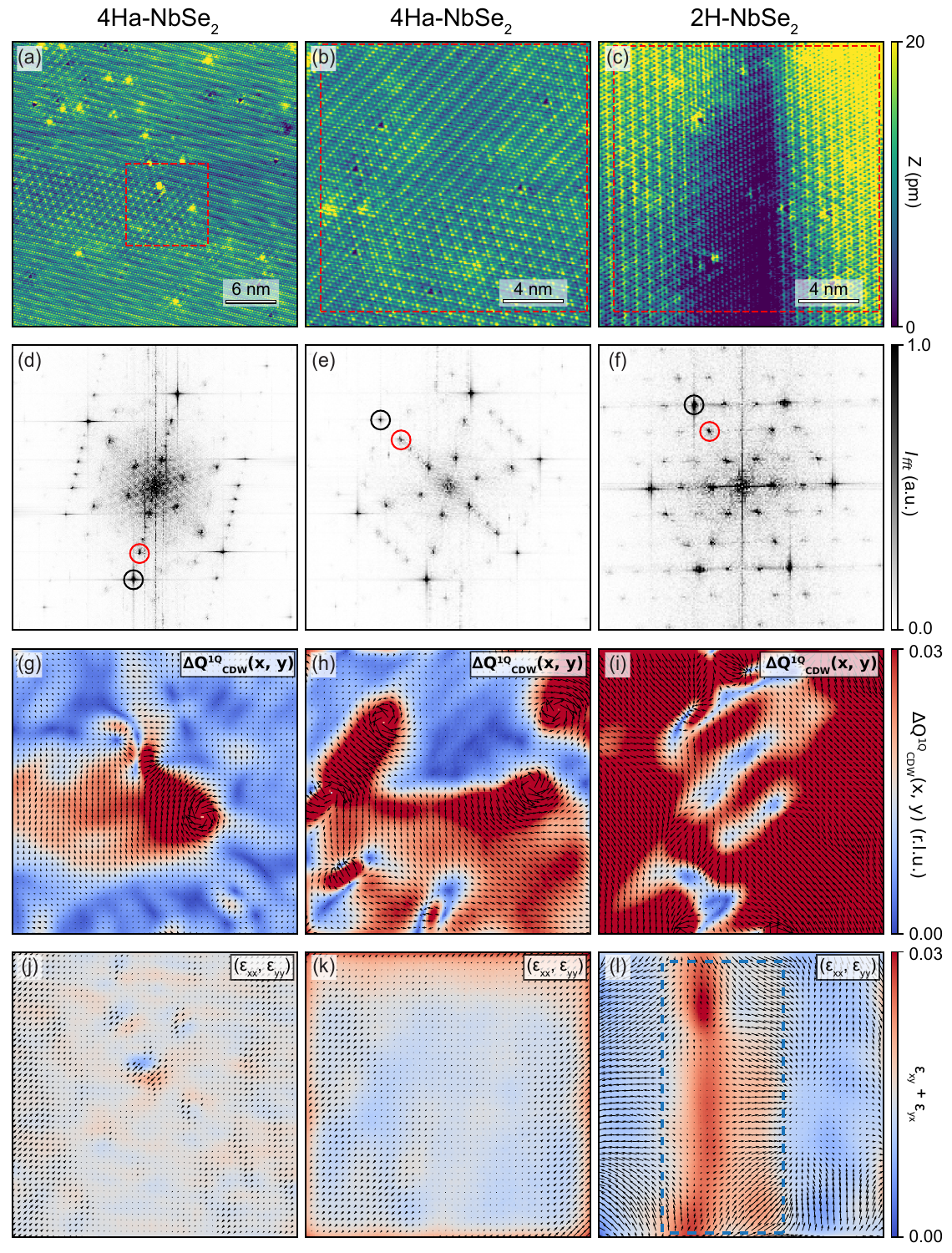}
    \caption{STM topographic images of regions showing the coexistence of the 1Q$_\mathrm{C}$ and 3Q$_\mathrm{I}$ charge-density-wave (CDW) phases in (a,b) 4Ha-NbSe$_2$ and (c) 2H-NbSe$_2$. The red dashed box marks the region where the LWM algorithm is applied, i.e. (g-l) are all the red dashed regions marked in the corresponding topographies. (d-f) Corresponding Fourier transforms of the images shown in the upper panels. Black and red circles mark the Bragg and CDW peaks used for the LWM algorithm. (g-i) $\Delta\bQ^{1Q}_{\mathrm{CDW}}(x, y)$ The background is the magnitude of the vector. (j-l) The vector field plots are vectors of the uniaxial strain ($\beps_{xx}, \beps_{yy}$) while the color background is the shear strain intensity $\beps_{xy} + \beps_{yx}$. The dashed blue box marks the region with large strain. (Measurement parameters for (a): $I_\mathrm{t}=80$~pA, $V_\mathrm{b}=100$~mV, $T=4.2$~K; (b): $I_\mathrm{t}=500$~pA, $V_\mathrm{b}=100$~mV, $T=4.2$~K; (c): $I_\mathrm{t}=90$~pA, $V_\mathrm{b}=1.4$~V, $T=3.7$~K.)}
    \label{fig:Experimental Data}
\end{figure*}

Here, we apply the LWM algorithm to experimental data to demonstrate a setting in which its high local wave vector resolution is particularly useful. We analyze constant-current STM topographic images acquired from single crystals of the transition-metal dichalcogenide NbSe$_2$ with two different stacking configurations, $\NbSeFourHa$ and $\NbSeTwoH$, whose electronic properties are strongly influenced by their stacking order. The measurements were performed in the STM laboratory at the Donostia International Physics Center and were previously reported in Ref.\,\cite{Sajan2026}, where the physical implications of the CDW phases in $\NbSeFourHa$ and $\NbSeTwoH$ were discussed in detail. Here, we use these data as a real experimental benchmark for the LWM algorithm. 

As shown in Ref.\,\cite{Sajan2026}, the conventional 2H polytype is dominated by an incommensurate triangular charge-density-wave (CDW) phase, whereas 4Ha-NbSe$_2$ hosts two competing CDW states: a unidirectional commensurate $1Q_\mathrm{C}$ phase with ordering wave vector $2/7\,\mathbf{Q}^{4\mathrm{Ha}}_{\mathrm{BRG}}$ and an incommensurate triangular $3Q_\mathrm{IC}$ phase with ordering wave vector close to $(2/7 + \delta)\,\mathbf{Q}^{4\mathrm{Ha}}_{\mathrm{BRG}}$. A unidirectional $1Q$ CDW can also appear in the 2H polytype, where it is strongly correlated with local strain \cite{Soumyanarayanan2013}. These differences motivate the questions of whether the LWM algorithm can resolve the small CDW wave vector differences associated with these phases and whether the formation of the $1Q_\mathrm{C}$ phase in 4Ha-NbSe$_2$ is driven by local lattice strain. Addressing these questions allows us to determine whether the frequent appearance of the $1Q_\mathrm{C}$ phase is primarily associated with the electronic structure or with geometric distortions of the lattice. Because the wave vector difference between the $1Q_\mathrm{C}$ and $3Q_\mathrm{IC}$ regions is small, the LWM algorithm provides better local resolution of the two phases than conventional local Fourier analysis. 

Figure\,\ref{fig:Experimental Data} shows STM topographic images of 4Ha-NbSe$_2$ [(a),(b)] and 2H-NbSe$_2$ [(c)], in which the coexistence of $1Q$ and $3Q$ phases is observed in both $\NbSeFourHa$ and $\NbSeTwoH$. The corresponding Fourier transforms are shown in Fig.\,\ref{fig:Experimental Data}(d--f) and exhibit $1Q_\mathrm{C}$ CDW peaks along one crystallographic direction, marked by red circles, as well as $3Q_\mathrm{IC}$ peaks along all three symmetry-equivalent crystallographic directions. Although the $1Q$ CDWs on the 4Ha and 2H surfaces exhibit similar wave vectors in the FFTs, the stripe phase in the 2H system is localized within a depressed region, whereas the 4Ha surface is atomically flat, with no detectable surface buckling. This contrast suggests that strain may drive the $1Q$ phase in the 2H system but not in the 4Ha system. 

To quantify the CDW incommensurability and its relationship to strain, we use the LWM algorithm to determine the local CDW and Bragg vectors along the $1Q_\mathrm{C}$ direction. We first obtain maps of the local Bragg vector, $\mathbf{Q}^{1Q}_{\mathrm{BRG}}(x,y)$, and the local CDW wave vector, $\mathbf{Q}^{1Q}_{\mathrm{CDW}}(x,y)$; the corresponding peaks in the FFT are marked by black and red circles, respectively. It is worth noting that we picked up the secondary $\frac{5}{7}$ peak rather than the primary $\frac{2}{7}$ to keep away from the central noise region of FFT and to obtain more wave front density to reduce the noise. We then define the empirical Bragg vector of the field of view (FOV), denoted by $\bar{\mathbf{G}}$, as the spatial average of $\mathbf{Q}^{1Q}_{\mathrm{BRG}}(x,y)$: \begin{equation} 
\bar{\mathbf{G}} = \left\langle \mathbf{Q}^{1Q}_{\mathrm{BRG}}(x,y) \right\rangle_{\mathrm{FOV}}. \end{equation} 
The local CDW incommensurability is defined as 
\begin{equation} \Delta\mathbf{Q}^{1Q}_{\mathrm{CDW}}(x,y) = \mathbf{Q}^{1Q}_{\mathrm{CDW}}(x,y) - \frac{5}{7}\bar{\mathbf{G}}, 
\end{equation} while the lattice strain is calculated from the Bragg-vector map $\mathbf{Q}^{1Q}_{\mathrm{BRG}}(x,y)$ using Eq.\,(\ref{Eq. Bragg Strain Relation}). As mentiioned before, the subtracted eavevector is the secondary peak of the CDW, $\frac{5}{7}\bar{\mathbf{G}} = -2/7\bar{\mathbf{G}} + \bar{\mathbf{G}}$. The vector-field plots of $\Delta\mathbf{Q}^{1Q}_{\mathrm{CDW}}(x,y)$ and the corresponding strain maps are shown in Figs.\,\ref{fig:Experimental Data}(g)--(i) and Figs.\,\ref{fig:Experimental Data}(j)--(l), respectively. The LWM algorithm clearly resolves the local CDW incommensurability, highlighted by the red regions in Figs.\,\ref{fig:Experimental Data}(g)--(i). CDW vortices also appear at the domain boundaries, where they accommodate the wavefront mismatch between the $1Q$ and $3Q$ regions, as discussed in our previous work \cite{Sajan2026}. 
This is analogous to the analysis of artificially generated vortices in the simulated data presented in Sec.\,\ref{subsection: Topological Defects} and Fig.\,\ref{Fig: Algorithm Result}, in which the phase-gradient field captures the vortex associated with a dislocation; here, the corresponding vortices are resolved by $\Delta\mathbf{Q}^{1Q}_{\mathrm{CDW}}(x,y)$. 
In contrast to the large variations in CDW incommensurability, the lattice strain in $\NbSeFourHa$ is nearly uniform across the FOV, both in the uniaxial-strain component represented by the vector field and in the shear-strain component shown by the color background, Figs.\,\ref{fig:Experimental Data}(j,k). This indicates that the CDW phase separation in $\NbSeFourHa$ is not driven by local lattice strain. By comparison, the strain map for $\NbSeTwoH$ (Fig.\,\ref{fig:Experimental Data}(l)) exhibits a clear stripe-like variation, marked by the dashed blue box, indicating that the phase-separation mechanisms differ between the 2H and 4Ha systems.

\section{Conclusion}

Measuring local strain and wave phases is increasingly important in STM studies of the relationship between electronic order and lattice structure. As shown by Eqs.\,(\ref{Eq. Bragg Strain Relation}) and (\ref{eq: Local Wave Vector and Phase Relation}), strain, local wave vector, and phase gradient are closely related quantities. Accurate measurement of any one of them can therefore provide access to the others. Traditionally, this problem has been addressed using the Lawler--Fujita algorithm, which extracts the phase of a periodic modulation and uses it to correct STM image drift and infer strain. However, this approach can be limited by failures of phase unwrapping and by noise amplification when the displacement field is differentiated. A complementary approach is to measure the local wave vector directly, but straightforward local Fourier-transform methods are limited by the tradeoff between real-space locality and resolution in Fourier space.

In this work, we developed the LWM algorithm, a real-space method that determines local wave vectors by measuring wavefront orientations with the Hough transform. We showed that this approach enables the local deformation gradient, and therefore the strain, to be obtained without first reconstructing a globally continuous phase field. We benchmarked the algorithm using synthetic STM-like images with known distortions and demonstrated that it can recover weak strain fields with high accuracy. We also showed that the method remains effective near topological defects, where phase unwrapping fails but the local phase gradient remains a well-defined and physically meaningful quantity. Finally, by applying the algorithm to experimental STM data from transition-metal dichalcogenides, we demonstrated that it can resolve small CDW wave vector incommensurabilities and distinguish them from lattice strain. Overall, the LWM algorithm provides a practical tool for extracting local strain and wave vector information from STM images, enabling quantitative studies of the interplay between electronic order and lattice geometry.

\section*{Acknowledgement}
M.N.G. is supported by the Ramon y Cajal Fellowship RYC2021-031639-I funded by MCIN/AEI/10.13039/501100011033. M.M.U. and M.N.G. also acknowledge support from grant PID2023-153277NB-I00 from the same institution. M.M.U. acknowledges support by the ERC Starting grant LINKSPM (Grant \#758558). EHdSN acknowledges support by the U.S. Department of Energy, Office of Basic Energy Sciences, Division of Materials Sciences and Engineering under Award DE-SC0025594.

\bibliography{main}

\end{document}